\documentclass[preprint,showpacs,amsmath,amssymb,aps,floats,a4paper]{revtex4}
\usepackage{bm}
\usepackage{graphicx}
\usepackage{times}
\newcommand{\beq}{\begin{equation}}
\newcommand{\eeq}{\end{equation}}

\newcommand{\bfb}{\mbox{\boldmath $b$}}

\newcommand{\bfh}{\mbox{\boldmath $h$}}
\newcommand{\bfk}{\mbox{\boldmath $k$}}
\newcommand{\bfu}{\mbox{\boldmath $u$}}
\newcommand{\bfv}{\mbox{\boldmath $v$}}
\newcommand{\bfx}{\mbox{\boldmath $x$}}
\newcommand{\bfB}{\mbox{\boldmath $B$}}
\newcommand{\bfH}{\mbox{\boldmath $H$}}
\newcommand{\bfK}{\mbox{\boldmath $K$}}
\newcommand{\bfR}{\mbox{\boldmath $R$}}
\newcommand{\bfX}{\mbox{\boldmath $X$}}
\newcommand{\bfxi}{\mbox{\boldmath $\xi$}}
\newcommand{\ex}{\mbox{{\boldmath $e$}}_{1}}
\newcommand{\ey}{\mbox{{\boldmath $e$}}_{2}}
\newcommand{\ez}{\mbox{{\boldmath $e$}}_{3}}
\newcommand{\bfemf}{\mbox{\boldmath ${\cal E}$}}
\newcommand{\bnabla}{\mbox{\boldmath $\nabla$}}
\newcommand{\cross}{\mbox{\boldmath $\times$}}
\newcommand{\cendot}{\mbox{\boldmath $\cdot\,$}}
\begin{document}

\title{A nonperturbative quasilinear approach to the shear dynamo problem}
\author{S. Sridhar}
\affiliation{Raman Research Institute, Sadashivanagar, Bangalore 560 080, India}
\email{ssridhar@rri.res.in}
\author{Kandaswamy Subramanian}
\affiliation{IUCAA, Post bag 4, Ganeshkhind, Pune 411 007, India}
\email{kandu@iucaa.ernet.in}

\date{\today}

\begin{abstract}
We study large--scale dynamo action due to turbulence 
in the presence of a linear shear flow. Our treatment is quasilinear  
and equivalent to the standard `first order smoothing approximation'.
However it is non perturbative in the shear strength.
We first derive an integro--differential equation for the 
evolution of the mean magnetic field, by systematic use of 
the shearing coordinate transformation and the Galilean invariance
of the linear shear flow. We show that, for non helical turbulence,
the time evolution of the cross--shear
components of the mean field do not depend on any 
other components excepting themselves;
this is valid for any Galilean--invariant velocity field, independent of its dynamics. 
Hence, to all orders in the shear parameter, there is no shear--current type effect
for non helical turbulence in a linear shear flow, in quasilinear theory in the 
limit of zero resistivity.
We then develop a systematic approximation of the integro--differential equation
for the case when the mean magnetic field varies
slowly compared to the turbulence correlation time.
For non-helical turbulence, the resulting partial differential 
equations can again be solved by making a shearing coordinate
transformation in Fourier space. The resulting 
solutions are in the form of shearing waves, labeled by
the wavenumber in the sheared coordinates. These shearing waves
can grow at early and intermediate times but are expected
to decay in the long time limit.
\end{abstract}

\maketitle

\section{Introduction}

The origin of large--scale magnetic fields in astrophysical systems
from stars to galaxies is an issue of considerable interest.
The standard paradigm involves dynamo amplification of seed magnetic fields
due to turbulent flows which have helicity combined with shear.
Shear flows and turbulence are ubiquitous in astrophysical systems
although the turbulence in general may not be helical.
However the presence 
of shear by itself may open new
pathways to the operation of large--scale dynamos,
even if the turbulence lacks a coherent helicity
\cite{BRRK08,Yousef08,PKB,RK03,Schek08}.
The evidence for such large--scale dynamo action
under the combined action of non helical
turbulence and background shear flow
comes mainly from several direct numerical simulations
\cite{BRRK08,Yousef08}.
How such a dynamo works is not yet clear.
One possibility is the shear--current effect \cite{RK03},
in which extra components of the mean electromotive force (EMF) arise due to shear,
which couple components of the mean magnetic field parallel and perpendicular to
the shear flow. However there is no convergence yet on whether the sign
of the relevant coupling term is such as to obtain a dynamo;
some analytic calculations \cite{RS06,RK06} and numerical experiments
\cite{BRRK08} find that the sign of the shear--current term is unfavorable
for dynamo action.

In an earlier paper \cite{SS09} (Paper I), we had outlined briefly
a quasilinear theory of dynamo action in a linear shear flow of an
incompressible fluid which has random
velocity fluctuations due either to freely decaying turbulence
or generated through external forcing.
Our analysis did not put any restrictions on the strength
of the shear, unlike earlier analytic work which treated
shear as a small perturbation.
We arrived at an integro--differential equation for 
the evolution of the mean magnetic field and argued
that the shear-current assisted dynamo is essentially absent
in quasilinear theory in the 
limit of zero resistivity.
In the present paper we give detailed derivations
of the main results of Paper I. We also extend our work further
by deriving differential equations for the mean field,
in the limit when the correlation time of the turbulence
is much smaller than the time-scale over which the mean field
varies. This allows us to solve for the mean field evolution
in terms of the velocity correlation functions. We can draw some general
conclusions on the shear dynamo independent of the exact 
velocity dynamics. In particular we note that the shear dynamo
can lead to transient growth of large-scale fields in the form of 
shearing waves, but these waves ultimately decay, even in the absence 
of microscopic diffusion.

In the section~II we formulate the shear dynamo problem. 
Our theory is `local' in character: In the laboratory frame we consider a 
background shear flow whose velocity is
unidirectional (along the $X_2$ axis) and varies linearly in 
an orthogonal direction (the $X_1$ axis). 
Section~III outlines a quasilinear theory of the shear dynamo. 
Systematic use of the shearing transformation allows us to develop a theory 
that is non perturbative in the strength of the background shear. However, we ignore 
the complications associated with nonlinear interactions, hence  magnetohydrodynamic (MHD)
turbulence and the small--scale dynamo; so our theory is quasilinear in nature, 
equivalent to the `first order smoothing approximation' (FOSA) \cite{dynam,BS05}.
The linear shear flow has a basic symmetry relating to measurements 
made by a special subset of all observers,
who may be called comoving observers. This symmetry is the invariance 
of the equations with respect to a group of transformations that is a subgroup 
of the full Galilean group. It may be referred to as `shear--restricted Galilean invariance', 
or Galilean invariance (GI). It should be noted that the laboratory frame and its set of 
comoving observers need not be inertial frames; in fact one of the main applications of 
GI is to the {\it shearing sheet} which is a rotating frame.
We introduce and explore the consequences of GI velocity 
fluctuations in section~IV. Such velocity fluctuations are not only compatible 
with the underlying symmetry 
of the problem, but they are expected to arise naturally. This has profound 
consequences for dynamo action, because the transport coefficients that define 
the mean EMF become spatially homogeneous in spite of the shear flow. 
The derivation of an integro-differential equation for the mean magnetic
field is given in section~V. We discuss a number of ways of approximating
this equation in section~VI, for slowly varying mean fields, all of
which lead to the same set of partial differential equations for the mean-field.
The mean field dynamics is further studied in section~VII, 
and section~VIII presents a discussion of the main results and 
the conclusions.

\section{The shear dynamo problem}
Let $(\ex,\ey,\ez)$ be the unit vectors of a Cartesian coordinate system in the laboratory frame,  
$\bfX = (X_1,X_2,X_3)$ the position vector, and $\tau$ the time. The fluid velocity is given by 
$(-2AX_1\ey + \bfv)$, where $A$ is the shear parameter 
(Oort's first constant) and $\bfv(\bfX, \tau)$ is a randomly fluctuating velocity field. The total magnetic field, $\bfB'(\bfX, \tau)$, obeys the induction equation:

\beq
\left(\frac{\partial}{\partial\tau} \;-\; 2AX_1\frac{\partial}{\partial X_2}\right)\bfB' \;+\; 2AB'_1\ey \;=\; 
\bnabla\cross\left(\bfv\cross\bfB'\right) \;+\; \eta\bnabla^2\bfB'\\[1em]
\label{indeqn}
\eeq
 
\noindent
The {\it shear dynamo problem} may be stated as follows: given some statistics of velocity fluctuations, what can be said about the magnetic field? More specific questions may be posed: does the combined action of the background shear and random velocities lead to the growth of a large--scale component of the magnetic field (i.e. a {\it turbulent dynamo})? In particular, is there turbulent dynamo action when the velocity fluctuations possess mirror--symmetry (i.e. when the velocity fluctuations are {\it non helical})? 

A common approach to the problem is through the theory of {\it mean--field electrodynamics}. Here,
the action of zero--mean velocity fluctuations ($\left<\bfv\right> = {\bf 0}$) on some seed magnetic field is assumed to produce a total magnetic field with a well--defined {\it mean--field} $(\bfB)$ and a {\it fluctuating--field} $(\bfb)$:

\beq
\bfB' \;=\; \bfB \;+\; \bfb\,,\qquad \left<\bfB'\right> \;=\; \bfB\,,\qquad \left<\bfb\right> \;=\; {\bf 0}
\label{reynolds}
\eeq

\noindent 
where $\left<\;\;\right>$ denotes ensemble averaging in the sense of Reynolds. Applying Reynolds averaging to the induction equation~(\ref{indeqn}), we obtain the following equations governing the dynamics of the mean and fluctuating magnetic fields:

\begin{eqnarray}
\left(\frac{\partial}{\partial\tau} \;-\; 2AX_1\frac{\partial}{\partial X_2}\right)\bfB \;+\; 2AB_1\ey &\;=\;& 
\bnabla\cross\bfemf \;+\; \eta\bnabla^2\bfB\label{meanindeqn}\\[2em] 
\left(\frac{\partial}{\partial\tau} \;-\; 2AX_1\frac{\partial}{\partial X_2}\right)\bfb \;+\; 2Ab_1\ey &\;=\;& 
\bnabla\cross\left(\bfv\cross\bfB\right) \;+\; \bnabla\cross\left(\bfv\cross\bfb - \bfemf\right) \;+\; \eta\bnabla^2\bfb\nonumber\\[1ex]
&&\label{flucindeqn}
\end{eqnarray}

\noindent
where $\bfemf = \left<\bfv\cross\bfb\right>$ is the mean EMF. The first step
toward solving the problem is to calculate $\bfemf$ and obtain a closed equation for the mean--field, 
$\bfB(\bfX, \tau)$. In the general case, it is necessary to specify the dynamics of $\bfv$ which
could be influenced by Lorentz forces due to both $\bfB$ and $\bfb$.

\section{Quasilinear theory}

To calculate the mean EMF we make some simplifying assumptions. We first make the {\it quasilinear} approximation in solving 
equation~(\ref{flucindeqn}) for $\bfb$ by dropping terms that 
that are quadratic in the fluctuations. Note that the dynamics of $\bfv$ is not prescribed; it does not imply
absence of velocity dynamics. For instance, the fluid can be acted upon by Lorentz forces due to the magnetic field, Coriolis force as in the case of the {\it shearing sheet}, or buoyancy in a convective flow. In this paper we will not specify any particular dynamics for the velocity field. We also drop the resistive term in the interests of simplicity of presentation. Setting $\eta=0$ may seem like a drastic step, but we would like to assure the reader that the theory can be reworked without this limitation and that our main conclusions carry through, even for $\eta \ne 0$. In particular we recover the results of this paper in the limit $\eta\to 0$. We note that the limit $\eta\to 0$ is also compatible with the 
physical situation in which the correlation times are small compared to the
eddy turn-over timescale; so our theory is applicable when the `first--order--smoothing--approximation' (FOSA) is valid. The fluctuating velocity field is assumed be incompressible 
$(\bnabla\cendot\bfv \;=\; 0)$. This restriction is not crucial and may be lifted without much difficulty.

\noindent
The quasilinear approximation is equivalent to neglecting the effects of magnetohydrodynamic 
turbulence and small--scale dynamo action, for the determination
of ${\cal E}$. With these assumptions, the equation for $\bfb$ we will solve is

\begin{eqnarray} 
\left(\frac{\partial}{\partial\tau} \;-\; 2AX_1\frac{\partial}{\partial X_2}\right)\bfb \;+\; 2Ab_1\ey &=& \bnabla\cross\left(\bfv\cross\bfB\right)\nonumber\\[1ex]
&\;=\;& \left(\bfB\cendot\bnabla\right)\bfv \;-\; \left(\bfv\cendot\bnabla\right)\bfB 
\label{flucindlineqn}
\end{eqnarray}

\subsection{The shearing coordinate transformation}

Equation~(\ref{flucindlineqn}) is inhomogeneous in the coordinate $X_1$. 
It is convenient to exchange spatial inhomogeneity for temporal inhomogeneity, so we 
get rid of the $\left(X_1{\partial/\partial X_2}\right)$ term through a shearing transformation to 
new spacetime variables:

\beq
x_1 = X_1\,,\qquad x_2 = X_2 + 2A\tau X_1\,,\qquad x_3 = X_3\,,\qquad t = \tau
\label{sheartr}
\eeq

\noindent
Then partial derivatives transform as

\beq
\frac{\partial}{\partial X_1} = \frac{\partial}{\partial x_1} + 
2At\frac{\partial}{\partial x_2}\,,\quad 
\frac{\partial}{\partial X_2}
= \frac{\partial}{\partial x_2}\,,\quad
\frac{\partial}{\partial X_3}
= \frac{\partial}{\partial x_3}\,,\quad
\frac{\partial}{\partial\tau} = \frac{\partial}{\partial t} + 
2Ax_1\frac{\partial}{\partial x_2} 
\label{partialsh}
\eeq

\noindent
We also define new variables, which are component--wise equal to the old variables: 

\beq
\bfH(\bfx, t) \;=\; \bfB(\bfX, \tau)\,,\qquad
\bfh(\bfx, t) \;=\; \bfb(\bfX, \tau)\,,\qquad
\bfu(\bfx, t) \;=\; \bfv(\bfX, \tau)
\label{newvar}
\eeq

\noindent
It is important to note that, just like the old variables, the new variables are expanded in the {\it fixed Cartesian basis of the laboratory frame}. For example, $\bfH = H_1\ex + H_2\ey + H_3\ez$, where $H_i(\bfx, t) =
B_i(\bfX, \tau)$, and similarly for the other variables. In the new variables, equation~(\ref{flucindlineqn}) becomes,

\beq
\frac{\partial\bfh}{\partial t} \;+\; 2Ah_1\ey \;=\; \left(\bfH\cendot\frac{\partial}{\partial\bfx}
+ 2AtH_1\frac{\partial}{\partial x_2}\right)\bfu \;-\; \left(\bfu\cendot\frac{\partial}{\partial\bfx} + 2Atu_1\frac{\partial}{\partial x_2}\right)\bfH
\label{newvareqn}
\eeq

\noindent
Equation~(\ref{newvareqn}) for $\bfh(\bfx, t)$ does not contain spatial derivatives  
of $\bfh$, so it can be integrated directly. We are interested in the particular solution which vanishes at 
$t=0$. The solutions for $h_1(\bfx, t)$ and $h_3(\bfx, t)$ are:

\begin{eqnarray}
h_1 &=& \int_0^t dt'\,u'_{1l}\left[H'_l + 2At'\delta_{l2}H'_1\right] \;-\;
\int_0^t dt'\,\left[u'_l + 2At'\delta_{l2}u'_1\right]H'_{1l}\label{h1soln}\\[3ex]
h_3 &=& \int_0^t dt'\,u'_{3l}\left[H'_l + 2At'\delta_{l2}H'_1\right] \;-\;
\int_0^t dt'\,\left[u'_l + 2At'\delta_{l2}u'_1\right]H'_{3l}\label{h3soln}
\end{eqnarray}

\noindent
where primes denote evaluation at spacetime point $(\bfx, t')$. We have also used notation $u_{ml} = 
(\partial u_m/\partial x_l)$ and $H_{ml} = (\partial H_m/\partial x_l)$.

The equation for $h_2(\bfx, t)$ involves  $h_1(\bfx, t)$; the solution is 

\beq
h_2 \;=\; \int_0^t dt'\,u'_{2l}\left[H'_l + 2At'\delta_{l2}H'_1\right] \;-\;
\int_0^t dt'\,\left[u'_l + 2At'\delta_{l2}u'_1\right]H'_{2l} \;-\;
2A\int_0^t dt'\, h'_1
\label{h2soln}
\eeq

\noindent 
We need to evaluate the integral

\beq
\int_0^t dt'\, h'_1 \;=\; \int_0^t dt'\int_0^{t'} dt^{''}\,u^{''}_{1l}\left[H^{''}_l + 2At^{''}\delta_{l2}H^{''}_1\right] \;-\;
\int_0^t dt'\int_0^{t'} dt^{''}\,\left[u^{''}_l + 2At^{''}\delta_{l2}u^{''}_1\right]H^{''}_{1l}
\eeq

\noindent
where the double--primes denote evaluation at spacetime point $(\bfx, t^{''})$. We now note that, 
for any function $f(\bfx, t)$, the double--time integral

\begin{eqnarray}
\int_0^t dt'\int_0^{t'} dt^{''}\,f(\bfx, t^{''}) &=& 
\int_0^t dt^{''}\,f(\bfx, t^{''})\int_{t^{''}}^{t}dt' \;=\;
\int_0^t dt^{''}\,(t - t^{''})\,f(\bfx, t^{''})\nonumber\\[2ex]
&=& \int_0^t dt'\,(t-t')\,f(\bfx, t')\nonumber
\end{eqnarray}

\noindent
reduces to a single--time integral, where in the last equality we have merely replaced 
the dummy integration variable $t^{''}$ by $t'$. Then 

\beq
\int_0^t dt'\, h'_1 \;=\; \int_0^t dt'\,(t-t')u'_{1l}\left[H'_l + 2At'\delta_{l2}H'_1\right] \;-\;
\int_0^t dt'\,(t-t')\left[u'_l + 2At'\delta_{l2}u'_1\right]H'_{1l}
\eeq

\noindent
can be used in equation~(\ref{h2soln}) to get an explicit solution for $h_2(\bfx, t)$. 
Combining equations~(\ref{h1soln}), (\ref{h3soln}) and (\ref{h2soln}) we can write  
$\bfh(\bfx, t)$ in component form as

\begin{eqnarray}
h_m(\bfx, t) \;\;=\;&& \int_0^t dt'\,\left[u'_{ml} \;-\; 2A(t-t')\delta_{m2}\,u'_{1l}\right]\left[H'_l \;+\; 2At'\delta_{l2}\,H'_1\right]
\nonumber\\[3ex]
&-&\int_0^t dt'\,\left[u'_l \;+\; 2At'\delta_{l2}\,u'_1\right]\left[H'_{ml} \;-\; 2A(t-t')\delta_{m2}\,H'_{1l}\right]
\label{hsoln}
\end{eqnarray}

\subsection{The mean EMF}

The expression in equation~(\ref{hsoln}) for $\bfh$ should be substituted in $\bfemf = \left<\bfv\cross\bfb\right> = \left<\bfu\cross\bfh\right>$. Following standard procedure, we allow $\left<\;\;\right>$ to act only on the velocity variables but not the mean field; symbolically, it is assumed that $\left<\bfu\bfu\bfH\right> = \left<\bfu\bfu\right>\bfH$. Interchanging the dummy indices $(l,m)$ in the last term of equation~(\ref{hsoln}), the mean EMF is given in component form as 

\begin{eqnarray}
{\cal E}_i(\bfx, t) \;\;=\;&& \epsilon_{ijm}\left<u_jh_m\right>\nonumber\\[3ex]
\;\;=\;&& \int_0^t dt'\;\,\left[\widehat{\alpha}_{il}(\bfx, t, t') \;-\; 2A(t-t')\widehat{\beta}_{il}(\bfx, t, t')
\right]\left[H'_l \;+\; 2At'\delta_{l2}\,H'_1\right]\nonumber\\[3ex]
&-&\int_0^t dt'\;\left[\,\widehat{\eta}_{iml}(\bfx, t, t') \;+\; 2At'\delta_{m2}\,\widehat{\eta}_{i1l}(\bfx, t, t')\right]\left[H'_{lm} \;-\; 2A(t-t')\delta_{l2}\,H'_{1m}\right]\nonumber\\[2ex]
&& \label{emf}
\end{eqnarray}

\noindent
where the {\it transport coefficients}, $(\widehat{\alpha}\,,\widehat{\beta}\,,\widehat{\eta}\,)$, are 
defined in terms of the $\bfu\bfu$ velocity correlators by

\begin{eqnarray}
\widehat{\alpha}_{il}(\bfx, t, t') &\;=\;& \epsilon_{ijm}\left<u_j(\bfx, t)\,u_{ml}(\bfx, t')\right>\nonumber\\[1ex]
\widehat{\beta}_{il}(\bfx, t, t') &\;=\;& \epsilon_{ij2}\left<u_j(\bfx, t)\,u_{1l}(\bfx, t')\right>\nonumber\\[1ex]
\widehat{\eta}_{iml}(\bfx, t, t') &\;=\;& \epsilon_{ijl}\left<u_j(\bfx, t)\,u_m(\bfx, t')\right>
\label{trcoeffs}
\end{eqnarray}

\noindent
To obtain more specific expressions for the transport coefficients, we need to provide information
on the $\bfu\bfu$ velocity correlators. However, it is physically more transparent to consider velocity statistics in terms of $\bfv\bfv$ velocity correlators, because this is referred to the laboratory frame instead of the sheared coordinates. By definition (eqn.~\ref{newvar}),

\beq
u_m(\bfx, t) \;=\; v_m(\bfX(\bfx,t), t)
\label{uvtr}
\eeq

\noindent
where

\beq
X_1 \;=\; x_1\,,\qquad X_2 \;=\; x_2 - 2Atx_1\,,\qquad X_3 \;=\; x_3\,,\qquad \tau \;=\; t
\label{invshtr}
\eeq

\noindent
is the inverse of the shearing transformation given in equation~(\ref{sheartr}). Using

\beq
\frac{\partial}{\partial x_l} \;=\; \frac{\partial}{\partial X_l} \;-\; 2A\tau\,\delta_{l1}\,\frac{\partial}{\partial X_2}
\label{invpartial}
\eeq

\noindent
the velocity gradient $u_{ml}$ can be written as

\beq
u_{ml} \;=\; \left(\frac{\partial}{\partial X_l} \;-\; 2A\tau\,\delta_{l1}\,\frac{\partial}{\partial X_2}\right)\,v_m \;=\; v_{ml} \;-\; 2A\tau\,\delta_{l1}\,v_{m2}
\label{uvgrad}
\eeq

\noindent
where $v_{ml} = (\partial v_m/\partial X_l)$. Then the transport coefficients are given in terms of the 
$\bfv\bfv$ velocity correlators by 

\begin{eqnarray}
\widehat{\alpha}_{il}(\bfx, t, t') &\;=\;& \epsilon_{ijm}\left[\left<v_j(\bfX, t)\,v_{ml}(\bfX', t')\right>
\;-\; 2At'\,\delta_{l1}\,\left<v_j(\bfX, t)\,v_{m2}(\bfX', t')\right>\right]\nonumber\\[1ex]
\widehat{\beta}_{il}(\bfx, t, t') &\;=\;& \epsilon_{ij2}\left[\left<v_j(\bfX, t)\,v_{1l}(\bfX', t')\right>
\;-\; 2At'\,\delta_{l1}\,\left<v_j(\bfX, t)\,v_{12}(\bfX', t')\right>\right]\nonumber\\[1ex]
\widehat{\eta}_{iml}(\bfx, t, t') &\;=\;& \epsilon_{ijl}\left<v_j(\bfX, t)\,v_m(\bfX', t')\right>
\label{trcoeffsvv}
\end{eqnarray}

\noindent
where $\bfX$ and $\bfX'$ are shorthand for
\beq
\bfX \;=\; \left(x_1\,,x_2 - 2Atx_1\,,x_3\right)\,,\qquad\bfX' \;=\; \left(x_1\,,x_2 - 2At'x_1\,,x_3\right)
\label{xxprime}
\eeq

Equation~(\ref{emf}), together with (\ref{trcoeffs}) or (\ref{trcoeffsvv}), gives the mean EMF in 
general form. $\bfX$ and $\bfX'$ can be thought of as the coordinates of the origin, at times 
$t$ and $t'$ respectively, of an observer {\it comoving} with the background shear flow. 
Therefore the transport coefficients depend only on the velocity correlators measured by such an observer 
at the origin of her coordinate system. This fact will have profound consequences for dynamo action, when we
consider G--invariant velocity correlators in the next section. Before discussing the Galilean invariance 
of the linear shear flow, we derive the form of the mean EMF for a special case, when the velocity
field is ``delta--correlated--in--time''.

\subsection{delta--correlated--in--time velocity correlator}

Although somewhat artificial, it is not uncommon to study dynamo action due to velocity fields 
whose correlation times are supposed so small that the two--point correlator taken between spacetime 
points $(\bfR, \tau)$ and $(\bfR', \tau')$ is assumed to be

\beq
\left<v_i(\bfR, \tau)\,v_j(\bfR', \tau')\right> \;=\; \delta(\tau - \tau')\,T_{ij}(\bfR, \bfR', \tau)
\label{deltacorr}
\eeq

\noindent
Incompressiblility implies that

\beq
\frac{\partial T_{ij}}{\partial R_i} \;=\; 0\,;\qquad
\frac{\partial T_{ij}}{\partial R'_j} \;=\; 0
\eeq

\noindent
We define

\beq
T_{ijl}(\bfR, \tau) \;=\; \left(\frac{\partial T_{ij}}{\partial R'_l}\right)_{\bfR'=\bfR}
\label{tijleqn}
\eeq

\noindent
The delta--function ensures that $\bfX$ and $\bfX'$ defined in equation~(\ref{xxprime}) are equal to 
each other. Then the velocity correlators

\begin{eqnarray}  
\left<v_i(\bfX,t)\,v_j(\bfX',t')\right> &\;=\;& \delta(t-t')\,T_{ij}(\bfX, \bfX, t) 
\nonumber\\[1ex]
\left<v_i(\bfX,t)\,v_{jl}(\bfX',t')\right> &\;=\;&  \delta(t-t')\,T_{ijl}(\bfX, t)
\label{deltavelcor}
\end{eqnarray}

\noindent
Substitute equation~(\ref{deltavelcor}) in equation~(\ref{trcoeffsvv}) for the transport coefficients; 

\begin{eqnarray}
\widehat{\alpha}_{il}(\bfx, t, t') &\;=\;& \delta(t-t')\,\epsilon_{ijm}\left[T_{jml} \;-\; 2At\,\delta_{l1}\,T_{jm2}\right]\nonumber\\[1ex]
\widehat{\beta}_{il}(\bfx, t, t') &\;=\;& \delta(t-t')\,\epsilon_{ij2}\left[T_{j1l} \;-\; 2At\,\delta_{l1}\,T_{j12}\right]\nonumber\\[1ex]
\widehat{\eta}_{iml}(\bfx, t, t') &\;=\;& \delta(t-t')\,\epsilon_{ijl}\,T_{jm}
\label{deltatrcoeff}
\end{eqnarray}

\noindent
and use these expressions in equation~(\ref{emf}). The delta--function ensures that the integrals over time 
can all be performed explicitly, so the mean EMF is

\beq
{\cal E}_i \;=\; \epsilon_{ijm}\left[T_{jml} \;-\; 2At\,\delta_{l1}\,T_{jm2}\right]
\left[H_l \;+\; 2At\,\delta_{l2}\,H_1\right] \;-\; 
\epsilon_{ijl}\left[T_{jm} \;+\; 2At\,\delta_{m2}\,T_{j1}\right]H_{lm}
\label{deltaemf}
\eeq

\noindent
It is useful to write the EMF in terms of the original variables and laboratory frame coordinates. To this end we 
transform

\beq
H_{lm} \;=\; \left(\frac{\partial}{\partial X_m} \;-\; 2A\tau\,\delta_{m1}\,\frac{\partial}{\partial X_2}\right)\,B_l \;=\; B_{lm} \;-\; 2A\tau\,\delta_{m1}\,B_{l2}
\label{HBgrad}
\eeq

\noindent where $B_{lm} = (\partial B_l/\partial X_m)$. Then the explicit dependence of ${\cal E}_i$ 
on the shear parameter $A$ cancels out, and the mean EMF assumes the simple form, 

\beq
{\cal E}_i \;=\; \epsilon_{ijm}\,T_{jml}\,B_l \;-\; \epsilon_{ijl}\,T_{jm}\,B_{lm}
\label{deltaemfsimple}
\eeq

\noindent
{\it which is identical to the familiar expression in the absence of background shear}. Therefore we conclude that, to obtain non trivial effects due to the shear flow, it is necessary to consider velocity correlators with non zero correlation times. Henceforth we shall consider the general case of finite velocity correlation times.  

\section{Galilean invariance}

The linear shear flow has a basic symmetry relating to measurements made by a special subset of 
all observers. We define a comoving observer as one whose velocity with respect to 
the laboratory frame is equal to the velocity of the background shear flow, and whose Cartesian coordinate
axes are aligned with those of the laboratory frame. A comoving observer can be labeled by the coordinates, 
$\bfxi = (\xi_1, \xi_2, \xi_3)$ with respect to the laboratory frame, of her origin at time $\tau=0$. 
Different labels identify different 
comoving observers and vice versa. As the labels run over all possible values, they exhaust the set of 
all comoving observers. The origin of the coordinate axes of a comoving observer translates 
with uniform velocity; its position with respect to the origin of the laboratory frame is given by 
\beq
\bfX_c(\tau) \;=\; \left(\xi_1\,, \xi_2 - 2A\tau \xi_1\,, \xi_3\right)
\label{orgvector}
\eeq
\noindent
An event with spacetime coordinates $(\bfX, \tau)$ in the laboratory frame has spacetime coordinates 
$(\tilde{\bfX}, \tilde{\tau})$ with respect to the comoving observer, given by
\beq
\tilde{\bfX} \;=\; \bfX \;-\; \bfX_c(\tau)\,,\qquad \tilde{\tau} \;=\; \tau - \tau_0
\label{coordtr}
\eeq
\noindent 
where the arbitrary constant $\tau_0$ allows for translation in time as well. 

Let $\left[\tilde{\bfB'}(\tilde{\bfX}, \tilde{\tau})\,, \tilde{\bfB}(\tilde{\bfX}, \tilde{\tau})\,, \tilde{\bfb}(\tilde{\bfX}, \tilde{\tau})\,,\tilde{\bfv}(\tilde{\bfX}, \tilde{\tau})\right]$ denote the 
total, the mean, the fluctuating magnetic fields and the fluctuating velocity field, respectively, as 
measured by the comoving observer. These are all equal to the respective quantities measured in the laboratory 
frame:

\beq
\left[\tilde{\bfB'}(\tilde{\bfX}, \tilde{\tau})\,, \tilde{\bfB}(\tilde{\bfX}, \tilde{\tau})\,,
\tilde{\bfb}(\tilde{\bfX}, \tilde{\tau})\,,\tilde{\bfv}(\tilde{\bfX}, \tilde{\tau})\right] \;=\; 
\left[\bfB'(\bfX, \tau)\,,\bfB(\bfX, \tau)\,,\bfb(\bfX, \tau)\,,\bfv(\bfX, \tau)\right]
\label{fields}
\eeq

\noindent
That this must be true may be understood as follows. Magnetic fields are invariant under non--relativistic 
boosts, so the total, mean and fluctuating magnetic fields must be the same in both frames.  
To see that the fluctuating velocity fields must be the same, we note that the total fluid 
velocity measured by the comoving observer is, by definition, 
equal to $\left(-2A\tilde{X_1}\ey + \tilde{\bfv}(\tilde{\bfX}, \tilde{\tau})\right)$. 
This must be equal to the difference between the 
velocity in the laboratory frame, $\left(-2AX_1\ey + \bfv(\bfX, \tau)\right)$, and $\left(-2A\xi_1\ey\right)$, 
which is the velocity of the comoving observer with respect to the laboratory frame. Using $\tilde{X} = X - \xi_1$, we see that $\tilde{\bfv}(\tilde{\bfX}, \tilde{\tau}) = \bfv(\bfX, \tau)$.

The {\it Galilean coordinate transformation} given in equation~(\ref{coordtr}) implies that partial derivatives are related 
through

\beq
\frac{\partial}{\partial\bfX}
\;=\; \frac{\partial}{\partial\tilde{\bfX}}\,,\qquad 
\frac{\partial}{\partial\tau} \;=\; \frac{\partial}{\partial\tilde{\tau}} \;+\; 
2A\xi_1\frac{\partial}{\partial\tilde{X_2}}
\label{partial}
\eeq

\noindent
Note that the combination $\left(\partial/\partial\tau - 2AX_1\partial/\partial X_2\right) = \left(\partial/\partial\tilde{\tau} - 2A\tilde{X_1}\partial/\partial \tilde{X_2}\right)$ is invariant in form. 
The other partial derivatives occurring in equations~(\ref{indeqn}), (\ref{meanindeqn}) and (\ref{flucindeqn}) are spatial derivatives which, by the second of equations~(\ref{partial}), are the same in both frames. Therefore equations~(\ref{indeqn}), (\ref{meanindeqn}) and (\ref{flucindeqn}) are invariant under the 
simultaneous transformations given in equations~(\ref{coordtr}) and (\ref{fields}). We note that this
symmetry property is actually invariance under a subset of the full ten--parameter Galilean group, parametrized
by the five quantities $\left(\xi_1, \xi_2, \xi_3, \tau_0, A\right)$; for brevity we will refer to this restricted symmetry as Galilean invariance, or simply GI. 

There is a fundamental difference between the coordinate transformations associated with Galilean invariance (equation~\ref{coordtr}) and the shearing transformation (equation~\ref{sheartr}). The former relates different 
comoving observers, whereas the latter describes a time--dependent distortion of the coordinates axes of one 
observer. Comparing equation~(\ref{partial}) with (\ref{partialsh}), we note that the relationship between old and new variables is homogeneous for the Galilean transformation, whereas it is  inhomogeneous for the shearing 
transformation.

It is important to note that the laboratory frame and its set of comoving observers need not be inertial frames. Indeed, one of the main applications of our theory is to the {\it shearing sheet} which is a rotating frame providing a local description of a differentially rotating disc; in addition to other forces, the velocity field is affected by the Coriolis force. The only requirement is that the magnetic field satisfies the induction equation~(\ref{indeqn}). 

\subsection{Galilean--invariant velocity correlators}

Naturally occurring velocity fields are Galilean--invariant, and this has a 
strong impact on the velocity statistics. We consider the $n$--point 
velocity correlator measured by the observer in the laboratory frame. Let this observer
correlate $v_{j_1}$ at spacetime location $(\bfR_1, \tau_1)$, with 
$v_{j_2}$ at spacetime location $(\bfR_2, \tau_2)$, and so on upto
$v_{j_n}$ at spacetime location $(\bfR_n, \tau_n)$. Now consider a comoving observer,
the position vector of whose origin is given by $\bfX_c(\tau)$ 
of equation~(\ref{orgvector}). An identical experiment performed by 
this observer must yield the same results, the measurements now made
at the spacetime points denoted by $\left(\bfR_1 +\bfX_c(\tau_1), \tau_1\right);
\left(\bfR_2 +\bfX_c(\tau_2), \tau_2\right);\ldots\,;\left(\bfR_n +\bfX_c(\tau_n), \tau_n\right)$.
If the velocity statistics is GI, the $n$--point velocity correlator must satisfy the condition

\beq
\left<v_{j_1}(\bfR_1, \tau_1)\,\ldots v_{j_n}(\bfR_n, \tau_n)\right> \;=\; 
\left<v_{j_1}(\bfR_1 + \bfX_c(\tau_1), \tau_1)\,\ldots v_{j_n}(\bfR_n + \bfX_c(\tau_n), \tau_n)\right> 
\label{ginvacorrn}
\eeq

\noindent
for all $(\bfR_1,\ldots\bfR_n\,; \tau_1,\ldots\tau_n\,; \bfxi)$. In quasilinear theory we 
require only the two--point velocity correlators, for which

\beq
\left<v_i(\bfR, \tau)\,v_j(\bfR', \tau')\right> \;=\; 
\left<v_i(\bfR + \bfX_c(\tau), \tau)\,v_j(\bfR' + \bfX_c(\tau'), \tau')\right> 
\label{ginvacorr}
\eeq

\noindent
for all $(\bfR, \bfR', \tau, \tau', \bfxi)$. We also need to work out the correlation between
velocities and their gradients:

\begin{eqnarray}
\left<v_i(\bfR, \tau)\,v_{jl}(\bfR', \tau')\right> &\;=\;&  
\frac{\partial}{\partial R'_l}\left<v_i(\bfR, \tau)\,v_j(\bfR', \tau')\right>\nonumber\\[2ex]
&\;=\;& \frac{\partial}{\partial R'_l}\left<v_i(\bfR + \bfX_c(\tau), \tau)\,v_j(\bfR' + \bfX_c(\tau'), \tau')\right>\nonumber\\[2ex]
&\;=\;& \left<v_i(\bfR + \bfX_c(\tau), \tau)\,v_{jl}(\bfR' + \bfX_c(\tau'), \tau')\right>
\label{ginvder}
\end{eqnarray}

\noindent
If we now set 

\beq
\bfR \;=\; \bfR' \;=\; {\bf 0}\,,\qquad \tau \;=\; t\,,\qquad \tau'\;=\; t'\,,\qquad
\left(\xi_1\,,\xi_2\,,\xi_3\right) \;=\; \left(x_1\,,x_2\,,x_3\right)
\label{setpar}
\eeq

\noindent
we will have

\beq
\bfX_c(\tau) \;=\; \left(x_1\,,x_2 - 2Atx_1\,,x_3\right)\,,\qquad\bfX_c(\tau') \;=\; 
\left(x_1\,,x_2 - 2At'x_1\,,x_3\right)
\label{setorgvec}
\eeq
 
\noindent
Comparing equation~(\ref{setorgvec}) with equation~(\ref{xxprime}), we see that $\bfX_c(\tau)$
and $\bfX_c(\tau')$ are equal to $\bfX$ and $\bfX'$, which are quantities that enter as arguments
in the velocity correlators of equations~(\ref{trcoeffsvv}) defining the transport coefficients. 
Hence, reading equations~(\ref{ginvacorr}) and (\ref{ginvder}) from right 
to left, the velocity correlators,

\begin{eqnarray}  
\left<v_i(\bfX,t)\,v_j(\bfX',t')\right> &\;=\;&  \left<v_i({\bf 0},t)\,v_j({\bf 0},t')\right> 
\;=\; R_{ij}(t, t')\nonumber\\[2ex]
\left<v_i(\bfX,t)\,v_{jl}(\bfX',t')\right> &\;=\;&  \left<v_i({\bf 0},t)\,v_{jl}({\bf 0},t')\right> 
\;=\; S_{ijl}(t, t')
\label{rsdef}
\end{eqnarray}

\noindent
are independent of space, and are given by the functions, $R_{ij}(t, t')$ and $S_{ijl}(t, t')$. Symmetry
and incompressiblity imply that $R_{ij}(t, t') = R_{ji}(t', t)$ and $S_{ijj}(t, t')=0\,$.
Note that the turbulence will, in general, be affected by the background shear and 
the velocity correlators will not be isotropic. In particular, $R_{ij}(t, t')$ will not be 
proportional to the unit tensor, $\delta_{ij}$.

\subsection{Galilean--invariant mean EMF}

The transport coefficients are completely determined by the form of the velocity correlator.
Using equations~(\ref{rsdef}) in equations~(\ref{trcoeffsvv}), we can see that the GI transport 
coefficients, 

\begin{eqnarray}
\widehat{\alpha}_{il}(t, t') &\;=\;& \epsilon_{ijm}\left[S_{jml}(t,t') \;-\; 2At'\,\delta_{l1}\,S_{jm2}(t,t')\right]\nonumber\\[1ex]
\widehat{\beta}_{il}(t, t') &\;=\;& \epsilon_{ij2}\left[S_{j1l}(t,t') \;-\; 2At'\,\delta_{l1}\,S_{j12}(t,t')\right]\nonumber\\[1ex]
\widehat{\eta}_{iml}(t, t') &\;=\;& \epsilon_{ijl}\,R_{jm}(t, t')
\label{trcoeffginv}
\end{eqnarray}

\noindent
are independent of space. Galilean invariance is the fundamental reason that the velocity correlators, hence the transport coefficients, are independent of space. The derivation given above is purely mathematical, relying on the basic freedom of choice of parameters made in equation~(\ref{setpar}), but we can also understand the results more physically. $\bfX$ and $\bfX'$, as given by equation~(\ref{xxprime}), can be thought of as the location of the origin of a comoving observer at times $t$ and $t'$, respectively. 
Thus when the observer correlates velocities at $\bfX = \bfX_c(t)$ and $\bfX' = \bfX_c(t')$, it will
be the same as correlating the velocities at her origin, but at different times. Then
GI implies that the velocity correlators  must be equal to those measured by {\it any} comoving observer at her origin at times $t$ and $t'$. In particular, this must be true for the observer in the laboratory frame, which explains equations~(\ref{rsdef}), consequently equations~(\ref{trcoeffginv}). 

We can derive an expression for the G--invariant mean EMF by using equations~(\ref{trcoeffginv})
for the transport coefficients in equation~(\ref{emf}). The integrands can be simplified as follows:

\begin{eqnarray}
\widehat{\alpha}_{il}(t, t')\left[H'_l + 2At'\delta_{l2}H'_1\right]
&\;=\;& \epsilon_{ijm}\left[S_{jml}(t,t') \;-\; 2At'\,\delta_{l1}\,S_{jm2}(t,t')\right]
\left[H'_l + 2At'\delta_{l2}H'_1\right]\nonumber\\[1ex]
&\;=\;& \epsilon_{ijm}S_{jml}(t,t')H'_l\nonumber\\[2ex]
\widehat{\beta}_{il}(t, t')\left[H'_l + 2At'\delta_{l2}H'_1\right]
&\;=\;& \epsilon_{ij2}\left[S_{j1l}(t,t') \;-\; 2At'\,\delta_{l1}\,S_{j12}(t,t')\right] 
\left[H'_l + 2At'\delta_{l2}H'_1\right]\nonumber\\[1ex]
&\;=\;& \epsilon_{ij2}S_{j1l}(t,t')H'_l\nonumber
\end{eqnarray}

\begin{eqnarray}
\left[\,\widehat{\eta}_{iml}(t, t') + 2At'\delta_{m2}\,\widehat{\eta}_{i1l}(t, t')\right]H'_{lm}
&\;=\;& \epsilon_{ijl}\left[R_{jm}(t, t') \;+\; 2At'\delta_{m2}\,R_{j1}(t, t')\right]H'_{lm}
\nonumber\\[2ex]
\left[\,\widehat{\eta}_{im2}(t, t') + 2At'\delta_{m2}\,\widehat{\eta}_{i12}(t, t')\right]H'_{1m}
&\;=\;& \epsilon_{ij2}\,\delta_{l1}\left[R_{jm}(t, t') \;+\; 2At'\delta_{m2}\,R_{j1}(t, t')\right]H'_{lm}
\nonumber
\end{eqnarray}

\noindent
Define
\begin{eqnarray}
C_{jml}(t,t') &\;=\;& S_{jml}(t,t') \;-\; 2A(t-t')\delta_{m2}\,S_{j1l}(t,t')\nonumber\\[2ex]
D_{jm}(t,t') &\;=\;& R_{jm}(t,t') \;+\; 2At'\delta_{m2}\,R_{j1}(t,t')
\label{cddef}
\end{eqnarray}

\noindent
The mean EMF can now be written compactly as

\beq
{\cal E}_i(\bfx, t)
\;=\; \epsilon_{ijm}\,\int_0^t dt'\;C_{jml}(t,t')H'_l
\;-\; \int_0^t dt'\,\left[\epsilon_{ijl} - 2A(t-t')\delta_{l1}\epsilon_{ij2}\right] D_{jm}(t, t')H'_{lm}
\label{emfcd}
\eeq

\noindent
where the $\bfx$ dependence of $\bfemf$ comes about only through the mean field, $\bfH(\bfx, t)$, and 
its spatial gradients, because the G--invariant transport coefficients are independent of $\bfx\,$.

\section{Mean--field induction equation}

Applying the shearing transformation given in equations~(\ref{sheartr}) and (\ref{partialsh}) to
the mean--field equation~(\ref{meanindeqn}), we see that the mean--field, $\bfH(\bfx, t)$, obeys

\beq
\frac{\partial H_i}{\partial t} \;+\; 2A\delta_{i2}H_1 \;=\;
\left(\bnabla\cross\bfemf\right)_i \;+\; \eta\bnabla^2 H_i
\label{hmeanindeqn}
\eeq

\noindent
where 

\beq
\left(\bnabla\right)_p \;\equiv\; \frac{\partial}{\partial X_p} \;=\; 
\frac{\partial}{\partial x_p} \;+\; 2At\,\delta_{p1}\frac{\partial}{\partial x_2}
\label{deltran}
\eeq

\noindent
It may be verified that equation~(\ref{hmeanindeqn}) preserves the condition $\bnabla\cendot\bfH =0\,$:

\beq
\bnabla\cendot\bfH \;\equiv\; \frac{\partial H_p}{\partial X_p} \;=\; H_{pp} \;+\; 2AtH_{12} \;=\; 0
\label{divcond}
\eeq

\noindent
We now use equations~(\ref{emfcd}) and (\ref{deltran}) to evaluate $\bnabla\cross\bfemf$.

\begin{eqnarray}
\left(\bnabla\cross\bfemf\right)_i &=& \epsilon_{ipq}\frac{\partial{\cal E}_q}{\partial X_p}
\;=\; \epsilon_{ipq}\left(\frac{\partial}{\partial x_p} \;+\; 2At\,\delta_{p1}\frac{\partial}{\partial x_2}
\right){\cal E}_q
\nonumber\\[3ex] 
&=& \epsilon_{ipq}\epsilon_{qjm}\,\int_0^t dt'\;C_{jml}(t,t')
\left[H'_{lp} + 2At\,\delta_{p1}H'_{l2}\right]\nonumber\\[2ex]
&&- \int_0^t dt'\, D_{jm}(t, t')\left[\epsilon_{ipq}\epsilon_{qjl} - 2A(t-t')\delta_{l1}\epsilon_{ipq}\epsilon_{qj2}
\right]\left[H'_{lmp} + 2At\,\delta_{p1}H'_{lm2}\right]
\nonumber
\end{eqnarray}

\noindent
Expanding $\epsilon_{ipq}\epsilon_{qjm} = \left(\delta_{ij}\,\delta_{mp} - \delta_{im}\,\delta_{jp}\right)$, 
the contribution from the $C$ term is

\beq
\left(\bnabla\cross\bfemf\right)^C_i \;=\; \int_0^t dt'\left[C_{ipl} - C_{pil}\right]
\left[H'_{lp} + 2At\delta_{p1}H'_{l2}\right]
\label{curlemfc}
\eeq

\noindent
Evaluating the $D$ term is a bit more involved. Again, we begin by expanding $\epsilon_{ipq}\epsilon_{qjl} = \left(\delta_{ij}\,\delta_{lp} - \delta_{il}\,\delta_{jp}\right)$. Then we get 

\begin{eqnarray}
\left(\bnabla\cross\bfemf\right)^D_i &=&
\int_0^t dt'\,D_{pm}\left\{H'_{ipm} + 2At\delta_{p1}H'_{i2m} - 
2A(t-t')\delta_{i2}\left[H'_{1pm} + 2At\delta_{p1}H'_{12m}\right]\right\}\nonumber\\[2ex]
&&- \int_0^t dt'\,D_{im}\left[H'_{ppm} + 2At'H'_{12m}\right]
\label{curlemfd}
\end{eqnarray}

\noindent
The second integral vanishes because the factor in $[\;]$ multiplying $D_{im}$ is zero: to see this,
differentiate the divergence--free condition of equation~(\ref{divcond}) 
with respect to $x_m$. Gathering together equations~(\ref{curlemfc}) 
and (\ref{curlemfd}), we have 

\begin{eqnarray}
\left(\bnabla\cross\bfemf\right)_i &=& \int_0^t dt'\left[C_{iml} - C_{mil}\right]
\left[H'_{lm} + 2At\delta_{m1}H'_{l2}\right] \;+\; \nonumber\\[2ex]
&+& \int_0^t dt'\,D_{jm}\left\{ H'_{ijm} + 2At\delta_{j1}H'_{i2m} -2A(t-t')\delta_{i2}\left[H'_{1jm} + 2At\delta_{j1}H'_{12m}\right]\right\} 
\label{curlemf}
\end{eqnarray}

\noindent
Thus the mean field $\bfH(\bfx, t)$ satisfies the mean--field 
induction equation,
\begin{eqnarray}
\frac{\partial H_i}{\partial t} &+& 2A\delta_{i2}H_1 =
\eta\bnabla^2 H_i +
\int_0^t dt'\left[C_{iml} - C_{mil}\right]
\left[H'_{lm} + 2At\delta_{m1}H'_{l2}\right] \;+\; \nonumber\\[2ex]
&+& \int_0^t dt'\,D_{jm}\left\{ H'_{ijm} + 2At\delta_{j1}H'_{i2m} -2A(t-t')\delta_{i2}\left[H'_{1jm} + 2At\delta_{j1}H'_{12m}\right]\right\} 
\label{Hinduction}
\end{eqnarray}
 
Equation~(\ref{Hinduction}) gives a closed set of 
integro--differential equations governing the dynamics of the mean--field, 
$\bfH(\bfx, t)$, valid for arbitrary values of $A$. Some of its important properties are: 

\begin{enumerate}

\item Only the part of $C_{iml}(t,t')$ that is antisymmetric in the indices $(i,m)$ 
contributes. 

\item The $D_{jm}(t,t')$ terms are such that $\left(\bnabla\cross\bfemf\right)_i$
involves only $H_i$ for $i=1$ and $i=3$, whereas $\left(\bnabla\cross\bfemf\right)_2$
depends on both  $H_2$ and $H_1$. This means that the mean--field induction 
equations~(\ref{Hinduction}) determining the time evolution 
of $H_1(\bfx, t)$ and $H_3(\bfx, t)$ are closed, whereas the equation for $H_2(\bfx, t)$ 
involves both $H_2(\bfx, t)$ and $H_1(\bfx, t)$. Thus $H_1(\bfx, t)$ (or $H_3(\bfx, t)$) can be computed by
using only the initial data $H_1(\bfx, 0)$ (or $H_3(\bfx, 0)$). The equation for $H_2$ 
involves both $H_2$ and $H_1$, and can then be solved. The implications for the original field, $\bfB(\bfX, \tau)$, can be read off, because it is 
equal to $\bfH(\bfx, t)$ component--wise (i.e $B_i(\bfX, \tau) = H_i(\bfx, t)$).
Thus, the $D_{jm}(t,t')$ terms do not couple either 
$B_1$ or $B_3$ with any other components, excepting themselves. In demonstrating this, we have 
not assumed that either the shear is small, or that $\bfH(\bfx, t)$ is such a slow function of 
time that it can be pulled out the time integral in equation~(\ref{curlemf}). 

\item When the turbulence is non helical,  $C_{iml}(t,t')=0$, but $D_{jm}(t,t')\neq 0$. In
this case, there is no shear--current type effect, in quasilinear theory in the limit of zero resistivity. 
This result should be compared with earlier work discussed in \cite{RK03,RS06,RK06}, where
there is explicit coupling of $B_2$ and $B_1$ in the evolution equation for $B_1$. A generalization of 
equation~(\ref{Hinduction}) to the case of non zero resistivity has been worked out in \cite{SSin09}. 
It is interesting to note that the corresponding generalization of $C_{iml}$ that appears in this case
need not vanish for non helical turbulence. However, it is expected to vanish in the formal limit of 
zero resistivity, consistent with our result given above.

\end{enumerate}

\section{The induction equation for a slowly varying mean--field}

\subsection{Mean EMF}

The mean EMF given in equation~(\ref{emfcd}) is a {\it functional} of $H_l$ and $H_{lm}$. 
When the mean--field is slowly varying compared to velocity correlation times, 
we expect to be able to approximate $\bfemf$ as a {\it function} of $H_l$ and $H_{lm}$. 
In this case, the mean--field induction equation
would reduce to a set of coupled partial differential equations, instead of the more 
formidable set of coupled integro--differential equations given by (\ref{hmeanindeqn}) 
and (\ref{curlemf}). Sheared coordinates are useful -- perhaps indispensable -- for 
calculations, but physical interpretation is simplest in the laboratory frame. Hence we derive an 
expression for the mean EMF in terms the original variables $B_l$ and $B_{lm}$. The result may 
be stated simply:

\begin{eqnarray}
{\cal E}_i &\;=\;& \alpha_{il}(\tau)\,B_l(\bfX, \tau) \;-\; \eta_{iml}(\tau)\,
\frac{\partial B_l}{\partial X_m}\nonumber\\[3ex]
\alpha_{il}(\tau) &\;=\;& \epsilon_{ijm}\,\int_0^{\tau} d\tau'\,
\left[C_{jml}(\tau,\tau') \;+\; 2A(\tau-\tau')\delta_{l1}C_{jm2}(\tau,\tau')\right]
\nonumber\\[3ex]
\eta_{iml}(\tau) &\;=\;& \epsilon_{ijl}\,\int_0^{\tau} d\tau'\,
\left[R_{jm}(\tau,\tau') \;-\; 2A(\tau-\tau')\delta_{m2}R_{j1}(\tau,\tau')\right]
\label{emfslowb}
\end{eqnarray}

\noindent
which is derived below by two different methods.

\subsubsection{Method I: use of a perturbative solution for $\bfH(\bfx, t')$}

Consider the mean--field equation~(\ref{hmeanindeqn}) when $\bfemf$ can be considered small.
We introduce an ordering parameter $\varepsilon\ll 1$ and consider $\bfemf$ to be $O(\varepsilon)$. 
Then a perturbative solution of equation~(\ref{hmeanindeqn}) in the $\eta\to 0$ limit
is

\beq
H_l(\bfx, t') \;=\;  H_l(\bfx, t)  \;+\; 2A(t-t')\delta_{l2}H_1(\bfx, t) \;+\;O(\varepsilon)
\label{pertsoln}
\eeq

\noindent
We can also consider perturbative solutions with non zero $\eta$, but using them in 
equation~(\ref{emfcd}) for $\bfemf$ would not be correct, because equation~(\ref{emfcd})
was derived in the limit $\eta\to 0$. We now use equation~(\ref{pertsoln}) in (\ref{emfcd}):

\begin{eqnarray}
{\cal E}_i(\bfx, t) &=& H_l\,\epsilon_{ijm}\int_0^t dt'\,C_{jml}(t,t') \;+\;
2AH_1\,\epsilon_{ijm}\int_0^t dt'\,(t-t')C_{jm2}(t,t')\nonumber\\[2ex]
&&\;-\; H_{lm}\,\epsilon_{ijl}\int_0^t dt'\,D_{jm}(t, t') \;+\; O(\varepsilon^2)
\label{emfslowtemp}
\end{eqnarray}

\noindent
Transform to the original field variables, using $H_l = B_l$ and $H_{lm} = B_{lm} - 
2At\delta_{m1}B_{l2}$, which is given in equation~(\ref{HBgrad}). The $C$ terms remain 
unaltered and can be seen to combine to equal $\alpha_{il}B_l$. Work out the $D$ term
using the expression for $D_{jm}$ given in equation~(\ref{cddef}):

\begin{eqnarray}
H_{lm}\,\int_0^t dt'\,D_{jm} &=& \left[B_{lm} - 
2At\delta_{m1}B_{l2}\right]\int_0^t dt'\,\left[R_{jm} + 2At'\delta_{m2}R_{j1}\right]\nonumber\\[2ex]
&=& B_{lm}\,\int_0^t dt'\,R_{jm} \;-\; 2AB_{l2}\,\int_0^t dt'\,(t-t')R_{j1}\nonumber
\end{eqnarray}

\noindent
Using the above result, and ignoring $O(\varepsilon^2)$ terms in equation~(\ref{emfslowtemp}), 
we obtain the result stated in  equation~(\ref{emfslowb}).

\subsubsection{Method II: Taylor expansion of $\bfB(\bfX', \tau' = t')$}

This is the standard approach, although not as short as the one given above. We express $\bfH(\bfx, t') = \bfB(\bfX', \tau' = t')$ and Taylor expand $\bfB$ inside the integral in equation~(\ref{emfcd}). As in equation~(\ref{xxprime}), 

\beq
\bfX = \left(x_1\,,x_2 - 2Atx_1\,,x_3\right)\,;\qquad \bfX' = \left(x_1\,,x_2 - 2At'x_1\,,x_3\right)
\nonumber
\eeq

\noindent
Writing $\bfX'=\bfX + 2A(t-t')x_1\ey$, we Taylor--expand:

\begin{eqnarray}
H'_l &\equiv\ & H_l(\bfx, t') \;=\; B_l(\bfX', t') \;=\; B_l(\bfX + 2A(t-t')x_1\ey, t')
\nonumber\\[2ex]
&=& B_l(\bfX, t) \;+\; 2A(t-t')x_1B_{l2} \;-\; (t-t')\frac{\partial B_l}{\partial t} 
\;+\; \ldots\nonumber
\end{eqnarray}

\noindent
We now use the mean--field induction equation~(\ref{meanindeqn}) to evaluate $(\partial B_l/\partial t)$.
As earlier we drop the contributions from $\left(\bnabla\cross\bfemf\right)$ and the $\eta$ term and get

\beq
\frac{\partial B_l}{\partial t} \;=\; 2Ax_1B_{l2} \;-\; 2A\delta_{l2}B_1 \;+\; \ldots
\label{bpertsoln}
\eeq

\noindent
Then
\begin{eqnarray}
H'_l &=& B_l(\bfX, t) \;+\; 2A(t-t')x_1B_{l2} \;-\; 
(t-t')\left[2Ax_1B_{l2} - 2A\delta_{l2}B_1\right] \;+\; \ldots\nonumber\\[2ex]
&=& B_l \;+\; 2A(t-t')\delta_{l2}B_1 \;+\; \ldots
\label{htaylor}
\end{eqnarray}

\noindent
Note that the inhomogeneous terms proportional to $x_1$ mutually cancel. It is clear, on 
physical grounds that they must, because the mean EMF given by equation~(\ref{emfcd}) is 
GI, and any valid approximation of a GI expression must preserve this symmetry. In 
particular, this  implies that transport coefficients cannot depend on $x_1$. We now use
equation~(\ref{htaylor}) inside the time integrals of (\ref{emfcd}). $B_l = B_l(\bfX, t)$ 
is a function of $(\bfx, t)$ and can be pulled out of the integrals over $t'$. Work out the 
$C$ and $D$ terms separately: 

\begin{eqnarray}
{\cal E}^C_i &=& \epsilon_{ijm}\,\int_0^t dt'\;C_{jml}(t,t')H'_l\nonumber\\[2ex]
&=& \epsilon_{ijm}\,\int_0^t dt'\;C_{jml}\left[B_l + 2A(t-t')\delta_{l2}B_1\right]\nonumber\\[2ex]
&=& \epsilon_{ijm}\,\int_0^t dt'\left[C_{jml}B_l + 2A(t-t')C_{jm2}B_1\right]\nonumber\\[2ex]
&=& \alpha_{il}\,B_l\nonumber
\end{eqnarray}

\noindent
To calculate the $D$ terms, we note that $H'_{lm} = (\partial H_l/\partial x_m)$. Since the
integral over $t'$ is performed at constant $\bfx$, the $(\partial /\partial x_m)$ can be 
pulled out of the integral:

\beq
{\cal E}^D_i \;=\; -\frac{\partial}{\partial x_m}\,\int_0^t dt'\,\left[\epsilon_{ijl} - 2A(t-t')\delta_{l1}\epsilon_{ij2}\right] D_{jm}(t, t')H'_l
\nonumber
\eeq

\noindent
Work out
 
\begin{eqnarray}
\left[\epsilon_{ijl} - 2A(t-t')\delta_{l1}\epsilon_{ij2}\right]\,H'_l &=&
\left[\epsilon_{ijl} - 2A(t-t')\delta_{l1}\epsilon_{ij2}\right]\,
\left[B_l + 2A(t-t')\delta_{l2}B_1\right]\nonumber\\[2ex]
&=& \epsilon_{ijl}B_l(\bfX, t)\nonumber
\end{eqnarray}

\noindent
Then

\beq
{\cal E}^D_i \;=\; - \epsilon_{ijl}\frac{\partial B_l}{\partial x_m}\,\int_0^t dt'\,D_{jm}(t, t')
\nonumber
\eeq

\noindent
The quantity

\beq
\frac{\partial B_l}{\partial x_m} \;=\; \left(\frac{\partial }{\partial X_m} \;-\;
2At\delta_{m1}\frac{\partial }{\partial X_2}\right)B_l \;=\;
B_{lm} \;-\; 2At\delta_{m1}B_{l2}\nonumber
\eeq

\noindent can be regarded as a function of $(\bfX, t)$ (or equivalently $(\bfx, t)$), and we are 
free to take it {\it inside} the $t'$ integral. When this is done and the expression 
for $D_{jm}$ given in equation~(\ref{cddef}) is used, we have

\begin{eqnarray}
{\cal E}^D_i &=&  - \epsilon_{ijl}\,\int_0^t dt'\,\left[B_{lm} - 2At\delta_{m1}B_{l2}\right]
\left[R_{jm} + 2At'\delta_{m2}R_{j1}\right]\nonumber\\[2ex]
&=&  - \epsilon_{ijl}B_{lm}\,\int_0^t dt'\,
\left[R_{jm} - 2A(t-t')\delta_{m2}R_{j1}\right]\nonumber\\[2ex]
&=& - \eta_{iml}B_{lm}
\end{eqnarray}

\subsection{Calculation of $\bnabla\cross\bfemf$}

We need to calculate $\bnabla\cross\bfemf$ for the mean EMF 
of equation~(\ref{emfslowb}). Work out the 
$\alpha$ and $\eta$ terms separately.

\begin{eqnarray}
\left(\bnabla\cross\bfemf\right)^{\alpha}_i &=& \epsilon_{ipq}\alpha_{il}B_{lp}
\;=\; B_{lp}\epsilon_{ipq}\epsilon_{qjm}\,\int_0^\tau d\tau'\,
\left[C_{jml} + 2A(\tau - \tau')\delta_{l1}C_{jm2}\right]\nonumber\\[4ex]
&=& B_{lm}\,\int_0^\tau d\tau'\,\left[C_{iml} + 2A(\tau - \tau')\delta_{l1}C_{im2}\right]\nonumber\\[2ex]
&&- \;B_{lj}\,\int_0^\tau d\tau'\,\left[C_{jil} + 2A(\tau - \tau')\delta_{l1}C_{ji2}\right]\nonumber\\[4ex]
&=& B_{lm}\,\int_0^\tau d\tau'\,\left\{C_{iml} - C_{mil} + 2A(\tau - \tau')\delta_{l1}
\left[C_{im2} - C_{mi2}\right]\right\}
\label{alphaterm}
\end{eqnarray}

\noindent 
Note that only the part of $C_{iml}$ that is antisymmetric in the indices $(i,m)$
contributes.

\begin{eqnarray}
\left(\bnabla\cross\bfemf\right)^{\eta}_i &=& - \epsilon_{ipq}\eta_{qml}B_{lpm}
\;=\; B_{lpm}\epsilon_{ipq}\epsilon_{qjl}\,\int_0^\tau d\tau'\,
\left[R_{jm} - 2A(\tau - \tau')\delta_{m2}R_{j1}\right]\nonumber\\[3ex]
&=& B_{ijm}\,\int_0^\tau d\tau'\,
\left[R_{jm} - 2A(\tau - \tau')\delta_{m2}R_{j1}\right]
\label{etaterm}
\end{eqnarray}

\noindent
where we have used $B_{ll} \equiv \bnabla\cendot\bfB = 0$. We note that equations~(\ref{alphaterm})
and (\ref{etaterm}) can also be derived directly from the expression for $\bnabla\cross\bfemf$, 
given in equation~(\ref{curlemf}). This is an interesting exercise as it allows us to 
formulate an alternate criteria on when the integral equation for
$\bfB$ can be approximated by differential equations.
We examine such an approximation further below.

\subsection{Approximating the integral equation directly}
It is convenient to work with the Fourier transform 
of $\bfH(\bfx, t)$:
\beq
\tilde{\bfH}(\bfk, t) \;=\; \int\,d^3x\,\bfH(\bfx, t)\exp{(-i\bfk\cendot\bfx)}
\label{fth1}
\eeq
We also define 
the vector $\bfK(\bfk, t) = (k_1 + 2At\,k_2, \,k_2, \,k_3)$ and $K^2 = \vert\bfK\vert^2 =
(k_1 + 2Atk_2)^2 + k_2^2 + k_3^2\,$: note that $\bfK\cdot\bfX = \bfk\cdot\bfx$. The magnetic 
field in the original variables, $\bfB(\bfX, \tau)$, can be recovered 
by using the shearing transformation, equation~(\ref{sheartr}), to write $(\bfx, t)$ 
in terms of the laboratory frame coordinates $(\bfX, \tau)$:
\begin{eqnarray}
\bfB(\bfX, \tau) &=& \bfH(\bfx, t) \;=\;
\int\,\frac{d^3k}{(2\pi)^3}\,\tilde{\bfH}(\bfk, t)\exp{(i\bfk\cendot\bfx)}\nonumber\\[3ex]
&=& \int\,\frac{d^3k}{(2\pi)^3}\,\tilde{\bfH}(\bfk, \tau)\exp{(i\bfK(\bfk, \tau)\cendot\bfX)}
\label{bmeansoln}
\end{eqnarray}
From equation~(\ref{Hinduction}), the Fourier transformed induction equation becomes
\begin{eqnarray}
\frac{\partial \tilde{H}_i}{\partial t} \;+\; 2A\delta_{i2}\tilde{H}_1 
&=&- \eta K^2 \tilde{H}_i \;+\; i \int_0^t dt'\left[C_{iml} - C_{mil}\right]
\left[\tilde{H}'_l k_m + 2At\delta_{m1}\tilde{H}'_l k_2\right] \; \nonumber\\[2ex]
&-& \int_0^t dt'\,D_{jm}\left\{ \tilde{H}'_i k_j k_m + 2At\delta_{j1}\tilde{H}'_i k_2k_m \right\} 
\nonumber\\[2ex]
&+& \int_0^t dt'\,D_{jm}\left\{ 2A(t-t')\delta_{i2}\left[\tilde{H}'_1 k_j k_m 
 + 2At\delta_{j1}\tilde{H}'_1 k_2 k_m \right]\right\} 
\label{curlemfk2}
\end{eqnarray}
Let us again simplify the integrals corresponding to the $C$ term, say $T^C$ 
and $D$ term, say $T^D$, separately. Using the definition of $\bfK(\bfk, t)$, the $C$ term
simplifies to
\begin{equation}
T^C_i \;=\; i K_m(\bfk, t) \int_0^t dt'\left[C_{iml} - C_{mil}\right]\tilde{H}'_l
\label{cterm}
\end{equation}
We now assume that the mean field is slowly varying compared to the 
correlation time $\tau_c$ of the turbulence
and Taylor expand $\tilde{H}_l(\bfk,t')$ about $t$ (this assumption
can later be checked for its self-consistency). We get
\begin{eqnarray}
\tilde{H}_l(\bfk,t') &=& \tilde{H}_l(\bfk,t) \;-\; (t-t') 
\frac{\partial \tilde{H}_l}{\partial t} \;+\; \ldots
\nonumber\\[2ex]
&=&
\left[\tilde{H}_l(\bfk,t) + 2A(t-t') \delta_{l2} \tilde{H}_1\right] - (t-t')\left[
\frac{\partial \tilde{H}_l}{\partial t} + 2 A \delta_{l2} \tilde{H}_1 \right] \;+\; \ldots
\label{taylorc}
\end{eqnarray}
where in the second line we have added and subtracted a term $2 A (t-t')\delta_{l2} \tilde{H}_1$.
Substituting this expansion in equation~(\ref{cterm}), the $C$-term becomes 
\begin{eqnarray}
T^C_i &=& iK_m(\bfk, t) \tilde{H}_l \int_0^t dt'
\left\{C_{iml} - C_{mil} + 2A(t - t')\delta_{l1}
\left[C_{im2} - C_{mi2}\right]\right\} 
\nonumber\\[2ex]
&-& iK_m(\bfk, t)\left[
\frac{\partial \tilde{H}_l}{\partial t} + 2 A \delta_{l2} \tilde{H}_1 \right]
\int_0^t dt' (t-t') [C_{iml} - C_{mil}]
\label{cterm2}
\end{eqnarray}

Now consider the $D$-terms. Again using the definition of $\bfK(\bfk, t)$ and $D_{jm} = R_{jm} + 2A t'\delta_{m2}R_{j1}$, we can simplify this to
\begin{equation}
T^D_i \;=\; -K_j K_m \int_0^t dt'\,[\tilde{H}'_i - 2A(t-t')\delta_{i2}\tilde{H}'_1]
[R_{jm} - 2A(t - t')\delta_{m2}R_{j1}]
\label{inducsim}
\end{equation}
Again assume that the mean field is slowly varying compared to the correlation time $\tau_c$ of the turbulence and Taylor expand $\tilde{H}_l(\bfk,t')$ about $t$. To first order in $(t-t')$, we have
\beq 
[\tilde{H}'_i - 2A(t-t')\delta_{i2}\tilde{H}'_1] \;=\;  \tilde{H}_i \;-\; 
(t-t')\left[\frac{\partial \tilde{H}_i}{\partial t} + 2 A \delta_{i2} \tilde{H}_1 \right] \;+\; \ldots
\nonumber
\eeq
Substituting this expansion in equation~(\ref{inducsim}) gives
\begin{eqnarray}
T^D_i  &=& -K_j K_m \tilde{H}_i \int_0^t dt'\,
[R_{jm} - 2A(t - t')\delta_{m2}R_{j1}]
\nonumber\\[2ex]
&+& K_j K_m \left[
\frac{\partial \tilde{H}_i}{\partial t} + 2 A \delta_{i2} \tilde{H}_1 \right]
\int_0^t dt'\, (t-t')[R_{jm} - 2A(t - t')\delta_{m2}R_{j1}]
\label{dterm2}
\end{eqnarray}

The expressions for $T^C_i$ and $T^D_i$ given in equations~(\ref{cterm2}) and (\ref{dterm2})
can be simplified. In both equations, the second terms are proportional to the LHS of the 
induction equation~(\ref{curlemfk2}). As before we ignore microscopic diffusion and write
\beq
\frac{\partial \tilde{H}_i}{\partial t} \;+\; 2 A \delta_{i2} \tilde{H}_1 \;\simeq\; 
T^C_i \;+\; T^D_i
\nonumber
\eeq
Then equations~(\ref{cterm2}) and (\ref{dterm2}) can be written as,
\begin{eqnarray}
T^C_i &=& iK_m(\bfk, t) \tilde{H}_l \int_0^t dt'
\left\{C_{iml} - C_{mil} + 2A(t - t')\delta_{l1}
\left[C_{im2} - C_{mi2}\right]\right\} \nonumber\\[2ex]
&-& iK_m(\bfk, t)\left[T^C_l + T^D_l\right]
\int_0^t dt' (t-t') [C_{iml} - C_{mil}]\nonumber\\[4ex]
T^D_i &=& -K_j K_m \tilde{H}_i \int_0^t dt'\,
[R_{jm} - 2A(t - t')\delta_{m2}R_{j1}]
\nonumber\\
&+& K_j K_m \left[T^C_i + T^D_i\right]
\int_0^t dt'\, (t-t')[R_{jm} - 2A(t - t')\delta_{m2}R_{j1}]
\label{cdterms}
\end{eqnarray}
When these equations are added together, they result in a set of three coupled linear equations 
for the unknown quantities $\left[T^C_1 + T^D_1\right]\,$, $\left[T^C_2 + T^D_2\right]\,$ and 
$\left[T^C_3 + T^D_3\right]\,$. It is straightforward to solve this system of equations, but the 
solutions assume a form which is needlessly complicated for our purposes. We are interested in the 
limit of short velocity correlations times, $\tau_c\,$. In this case both $T^C_i$ and $T^D_i$ are well approximated by their respective first terms:
\begin{eqnarray}
T^C_i &=& iK_m(\bfk, t) \tilde{H}_l \int_0^t dt'
\left\{C_{iml} - C_{mil} + 2A(t - t')\delta_{l1}
\left[C_{im2} - C_{mi2}\right]\right\} \nonumber\\[3ex]
T^D_i &=& -K_j K_m \tilde{H}_i \int_0^t dt'\,
[R_{jm} - 2A(t - t')\delta_{m2}R_{j1}]
\label{cdtermsfinal}
\end{eqnarray}
These are exactly the Fourier transforms of equation~(\ref{alphaterm}) for
$(\bnabla\cross\bfemf)^{\alpha}_i$, and equation~(\ref{etaterm}) for
$(\bnabla\cross\bfemf)^{\eta}_i$. 

We now state the conditions under which the approximations given in 
equations~(\ref{cdtermsfinal}) are valid. Let us define the quantitites $\alpha_0$ and $\eta_{\rm turb}$ as typical values of the time integrals of the velocity correlators, $S_{jml}$ and $R_{jm}$, respectively (for homogeneous and isotropic turbulence, $\alpha_0$ is of order the magnitude of the usual $\alpha$-effect, and $\eta_{\rm turb}$ would be comparable to the magnitude of the usual turbulent diffusion coefficient). For any wavenumber, $K$,
we can define time scales, $t_\alpha = (K\alpha_0)^{-1}$ and $t_\eta = (K^2\eta_{\rm turb})^{-1}$, 
associated with $\alpha_0$ and $\eta_{\rm turb}$. When $\tau_c$ is small enough such that
\begin{eqnarray}
\tau_c &\ll& t_\alpha\,,t_\eta\,;\qquad\mbox{$A\tau_c\ll 1$}\nonumber\\[1ex]
A\tau^2_c &\ll& t_\alpha\,,t_\eta\,;\qquad\mbox{$A\tau_c \geq 1$}
\label{smallcorrtime}
\end{eqnarray}
then both $T^C_i$ and $T^D_i$ are well approximated by their respective first terms, as
given in equations~(\ref{cdtermsfinal}).
The time scales, $t_\alpha = (K\alpha_0)^{-1}$ and $t_\eta = (K^2\eta_{\rm turb})^{-1}$, depend 
on the spatial scale, $K^{-1}$, which is a time--dependent quantity for $k_2\neq 0$; at late times, 
$K \sim \vert 2Atk_2\vert$ and this makes the quantities $t_\alpha$ and $t_\eta$ decreasing functions of 
time. With this fact taken into account, the inequalities given in equation~(\ref{smallcorrtime})
translate into upper limits on the time over which the expressions in equation~(\ref{cdtermsfinal}) serve 
as good approximations to $T^C_i$ and $T^D_i$.

\subsection{Mean--field induction equation}

We gather together here the results obtained in this section. When the mean--field is slowly varying, it 
satisfies the following partial differential equation:

\beq
\left(\frac{\partial}{\partial\tau} \;-\; 2AX_1\frac{\partial}{\partial X_2}\right)B_i 
\;+\; 2A\delta_{i2}B_1 \;=\; 
\tilde{\alpha}_{imj}(\tau)\frac{\partial B_j}{\partial X_m}
\;+\; \tilde{\eta}_{jm}(\tau)\frac{\partial^2 B_i}{\partial X_j\partial X_m}
\;+\; \eta\bnabla^2 B_i
\label{meanindeqnfinal}
\eeq

\noindent
where  

\begin{eqnarray}
\tilde{\alpha}_{imj}(\tau)  &\;=\;&
\int_0^\tau d\tau'\,\left\{C_{imj} - C_{mij} \;+\; 2A(\tau - \tau')\delta_{j1}
\left[C_{im2} - C_{mi2}\right]\right\}\nonumber\\[4ex]
\tilde{\eta}_{jm}(\tau) &\;=\;& \frac{1}{2}\,\int_0^\tau d\tau'\,
\left\{R_{jm} + R_{mj} \;-\; 2A(\tau - \tau')\left[\delta_{m2}R_{j1} + \delta_{j2}R_{m1}\right]\right\}
\label{curlemffinal}\\[2ex]\nonumber
\end{eqnarray}

\noindent
In the above integrals $C_{imj} = C_{imj}(\tau,\tau')$, $R_{jm} = R_{jm}(\tau,\tau')$ etc. 
Some comments:

\begin{enumerate}

\item Note that $\tilde{\alpha}_{imj}$ is antisymmetric in the indices $(i, m)$, whereas
$\tilde{\eta}_{jm}$ is symmetric in the indices $(j, m)$.

\item $\tilde{\eta}_{jm}$ terms do not lead to coupling of any component of $\bfB$
with any other component. 
\end{enumerate}

\section{Mean--field dynamics for non helical velocity statistics}

When the velocity fluctuations are non helical, $S_{imj}(\tau, \tau') = 0$, so that both 
$C_{imj}(\tau, \tau')$ and $\tilde{\alpha}_{mj}(\tau)$ vanish (in specific models of the velocity 
dynamics we find that the generated velocity fluctuations are indeed non-helical, if the forcing 
is non helical even in the presence of shear). Then the evolution of the mean--field,
(over times when the inequalities of equations~\ref{smallcorrtime} are satisfied), is determined by 

\beq
\left(\frac{\partial}{\partial\tau} \;-\; 2AX_1\frac{\partial}{\partial X_2}\right)B_i 
\;+\; 2A\delta_{i2}B_1 \;=\; 
\tilde{\eta}_{jm}(\tau)\frac{\partial^2 B_i}{\partial X_j\partial X_m} \;+\;
\eta\bnabla^2 B_i
\label{meanindeqnnh}
\eeq

\noindent
Note that $\tilde{\eta}_{jm}$ depends on the nature of the stirring and will, in general, be 
a function of time; this will be the case, say, for decaying turbulence. However, for statistically stationary
stirring, $\tilde{\eta}_{jm}$ will become time--independent, after an initial transient evolution. 

Equation~(\ref{meanindeqnnh}) is inhomogeneous in the spatial coordinates so, as before, we find it 
convenient to work with the new variable, $\bfH(\bfx, t)$, and transform equation~(\ref{meanindeqnnh}) 
to the shearing coordinates $(\bfx, t)$:

\beq
\frac{\partial H_i}{\partial t} \;+\; 2A\delta_{i2}H_1 \;=\;
\tilde{\eta}_{jm}(\tau)\frac{\partial^2 H_i}{\partial X_j\partial X_m} \;+\;
\eta\bnabla^2 H_i
\label{hmeaneqnnh}
\eeq

\noindent
where (see eqn.~\ref{deltran}) 

\beq
\frac{\partial}{\partial X_p} \;=\; 
\frac{\partial}{\partial x_p} \;+\; 2At\,\delta_{p1}\frac{\partial}{\partial x_2}\;;\qquad
\bnabla^2 \;=\; 
\left(\frac{\partial}{\partial x_p} \;+\; 2At\,\delta_{p1}\frac{\partial}{\partial x_2}\right)^2
\label{deltrannh}
\eeq 

\noindent
Equation~(\ref{hmeaneqnnh}) is homogeneous in $\bfx$ but not in $t$, so we take a 
spatial Fourier transform defined earlier in equation~(\ref{fth1}). Then $\tilde{\bfH}(\bfk, t)$ satisfies

\beq
\frac{\partial \tilde{H}_i}{\partial t} \;+\; 2A\delta_{i2}\tilde{H}_1 \;=\;
-\left[\tilde{\eta}_{jm}(t)\,K_jK_m \;+\; \eta\,K^2\right]\tilde{H}_i
\label{meanhfteqn}
\eeq

\noindent
where the vector $\bfK(\bfk, t) = (k_1 + 2At\,k_2, \,k_2, \,k_3)$ and $K^2 = \vert\bfK\vert^2 =
(k_1 + 2Atk_2)^2 + k_2^2 + k_3^2\,$, as before. It may be verified that this equation preserves the 
Fourier version of the divergence condition of equation~(\ref{divcond}), namely $\bfK\cendot\tilde{\bfH}(\bfk,t) = 0\,$. The solution is 

\begin{eqnarray}
\tilde{H}_1(\bfk, t) &=& \tilde{H}_1(\bfk,0)\,{\cal G}(\bfk,t)\nonumber\\[2ex]
\tilde{H}_2(\bfk, t) &=& \left[\tilde{H}_2(\bfk,0)
\;-\; 2At\,\tilde{H}_1(\bfk, 0)\right]{\cal G}(\bfk,t)\nonumber\\[2ex]
\tilde{H}_3(\bfk, t) &=& \tilde{H}_3(\bfk,0)\,{\cal G}(\bfk,t)
\label{hmeansoln}
\end{eqnarray}

\noindent
where $\tilde{\bfH}(\bfk, 0)$ are given initial conditions satisfying $\bfk\cendot\tilde{\bfH}(\bfk, 0) =0$,
ensuring that $\bfK\cendot\tilde{\bfH}(\bfk,t) = 0\,$. The Green's function, ${\cal G}(\bfk,t)$, is zero for $t<0$ and is defined for $t\geq 0$ by

\beq
{\cal G}(\bfk,t) \;=\; \exp{\left[-\int_0^t\, ds\left(\tilde{\eta}_{jm}(s)K_jK_m + \eta K^2\right)\right]}
\label{greenfn}
\eeq

\noindent
In the integrand, $K_j = k_j + 2As\delta_{j1}k_2$  should be regarded as a function of $\bfk$ and $s$, and the $s$--integral performed at fixed $\bfk$. Then ${\cal G}(\bfk,t)$ can be written as the product of 
a {\it microscopic} Green's function, ${\cal G}_\eta(\bfk,t)$,  and a {\it turbulent} Green's function, 
${\cal G}_{\rm t}(\bfk,t)\,$:

\begin{eqnarray}
{\cal G}(\bfk,t) &=& {\cal G}_\eta(\bfk,t)\cdot{\cal G}_{\rm t}(\bfk,t)\nonumber\\[4ex]
{\cal G}_\eta(\bfk,t) &=& \exp{\left[-\eta\left(k^2\,t + 2Ak_1k_2\,t^2
+ \frac{4}{3}A^2k_2^2\,t^3\right)\right]}\nonumber\\[4ex]
{\cal G}_{\rm t}(\bfk,t) &=& \exp{\left[-Q_{jm}(t)k_jk_m\right]}
\label{greenfnexp} 
\end{eqnarray}

\noindent
where the time--dependent symmetric matrix $Q_{jm}(t)$ is given by

\beq
Q_{jm}(t) \;=\; \int_0^t\,ds\left\{\tilde{\eta}_{jm}(s) + 2As\left[\delta_{j2}\,\tilde{\eta}_{1m}(s) + \delta_{m2}\,\tilde{\eta}_{j1}(s)\right]
+ 4A^2\delta_{j2}\,\delta_{m2}\,s^2\,\tilde{\eta}_{11}(s)\right\}
\label{qdef}
\eeq
in terms of time integrals of $\tilde{\eta}_{jm}(\tau)$, which are assumed to be known functions 
depending on the velocity correlators, $R_{jm}(\tau,\tau')$, as given in equation~(\ref{curlemffinal}). 

\noindent
The solution in the original variables, $\bfB(\bfX, \tau)$, can be recovered by using the shearing transformation, equation~(\ref{sheartr}), to write $(\bfx, t)$ in terms of the laboratory frame coordinates
$(\bfX, \tau)$ (see equation~\ref{bmeansoln}):
\begin{eqnarray}
\bfB(\bfX, \tau) &=& \bfH(\bfx, t) \;=\; 
\int\,\frac{d^3k}{(2\pi)^3}\,\tilde{\bfH}(\bfk, t)\exp{(i\bfk\cendot\bfx)}\nonumber\\[3ex]
&=& \int\,\frac{d^3k}{(2\pi)^3}\,\tilde{\bfH}(\bfk, \tau)\exp{(i\bfK(\bfk, \tau)\cendot\bfX)}
\label{bmeansoln2}
\end{eqnarray}

\noindent
Equivalently, the solution is given in component form as

\begin{eqnarray}
B_1(\bfX, \tau) &=& \int\,\frac{d^3k}{(2\pi)^3}\,\tilde{B}_1(\bfk,0)\,{\cal G}(\bfk, \tau)\,\exp{(i\bfK(\bfk, \tau)\cendot\bfX)}
\nonumber\\[4ex]
B_2(\bfX, \tau) &=& \int\,\frac{d^3k}{(2\pi)^3}\,\left[\tilde{B}_2(\bfk,0) - 2A\tau\tilde{B}_1(\bfk, 0)\right]
{\cal G}(\bfk, \tau)\,\exp{(i\bfK(\bfk, \tau)\cendot\bfX)}\nonumber\\[4ex]
B_3(\bfX, \tau) &=& \int\,\frac{d^3k}{(2\pi)^3}\,\tilde{B}_3(\bfk,0)\,{\cal G}(\bfk, \tau)\,\exp{(i\bfK(\bfk, \tau)\cendot\bfX)}
\label{bmeancompsoln}
\end{eqnarray}

\noindent
where we have written the initial condition, $\tilde{\bfH}(\bfk, 0) = \tilde{\bfB}(\bfk, 0)$, with
$\bfk\cendot\tilde{\bfB}(\bfk, 0) =0$. 

Some comments:

\begin{enumerate}

\item The above solution for $\bfB(\bfX, \tau)$ is a linear superposition of {\it shearing waves}, of the form $\exp{(i\bfK(\bfk, \tau)\cendot\bfX)} = \exp{\left[i(k_1 + 2A\tau k_2)X_1 + ik_2X_2 + ik_3X_3\right]}$, indexed by the triplet of numbers $(k_1, k_2, k_3)$. 

\item Whether the waves grow or decay depends on the time dependence of the Green's function, ${\cal G}(\bfk,\tau) = {\cal G}_\eta(\bfk,\tau)\cdot{\cal G}_{\rm t}(\bfk,\tau)$. The first term, ${\cal G}_\eta$,
is known explicitly and describes the ultimately decay of the shearing waves (on the long resistive timescale), 
although these could be transiently amplified. The second term, ${\cal G}_{\rm t}$, depends on the properties of  the time--dependent symmetric matrix $Q_{jm}(\tau)$.  Shearing waves can grow if $Q_{jm}(\tau)$ has at least one negative eigenvalue of large enough magnitude. To translate this requirement into an explicit statement on dynamo action requires developing a dynamical theory of the velocity correlators, $R_{jm}(\tau,\tau')$, because $Q_{jm}(\tau)$ depends on time integrals over $R_{jm}(\tau,\tau')$. 
\end{enumerate}

In specific cases it is possible that the velocity dynamics is such that
$\tilde{\eta}_{jm}(\tau)$ becomes independent of $\tau$, in the long time limit
(this is generic when steady forcing competes with dissipation). Taking the zero of time 
to be after this stationary state has been reached, we can do the $s$ integrals in 
equation~(\ref{qdef}) explicitly and write

\beq
Q_{jm}(t)k_jk_m \;=\; t \, (\tilde{\eta}_{jm}k_jk_m) 
+ 2At^2 \, (\tilde{\eta}_{1m}k_m k_2)
+ \frac{4}{3}A^2t^3 \, (\tilde{\eta}_{11} k_2^2)
\label{qcon}
\eeq
We can now make further statements on the dynamo growth using equation~(\ref{qcon}).
Note that the linear shear of the form that we have adopted is likely to lead to a non-zero $\tilde{\eta}_{12}$, but is not expected to couple the $X_3$ component with other
components, and thus we expect $\tilde{\eta}_{13}= \tilde{\eta}_{23} =0$. Then 
\beq
-Q_{jm}k_jk_m = -t \, 
\left[\tilde{\eta}_{11} k_1^2 + \tilde{\eta}_{22}k_2^2 + 
2\tilde{\eta}_{12}k_1k_2 + \tilde{\eta}_{33}k_3^2)\right]
-2A t^2 \, \left[\tilde{\eta}_{11}k_1 k_2 +\tilde{\eta}_{12}k_2^2\right]
-\frac{4}{3} A t^3 \, \tilde{\eta}_{11} k_2^2
\label{qspecific}
\eeq
The term linear in $t$ will dominate at early times while
the term proportional to $t^3$ will dominate
eventually. Thus at early times we need one of the eigenvalues of the
matrix
\[
\begin{pmatrix}
\tilde{\eta}_{11} & \tilde{\eta}_{12} & 0 \\ 
\tilde{\eta}_{12} & \tilde{\eta}_{22} & 0 \\
0 & 0 & \tilde{\eta}_{33}  
\end{pmatrix}
\label{mat}
\]
\noindent
to be negative for dynamo growth. These eigenvalues are 
\beq
\lambda_{\pm} = \frac{(\tilde{\eta}_{11} + \tilde{\eta}_{22})}{2}
\pm \frac{\vert \tilde{\eta}_{11} - \tilde{\eta}_{22}\vert}{2}
\left[ 1 + 4\frac{\tilde{\eta}_{12}^2}{(\tilde{\eta}_{11} - \tilde{\eta}_{22})^2}
\right]^{1/2};
\qquad
\lambda_3 = \tilde{\eta}_{33} 
\label{eigen}
\eeq
Nonzero values of $\tilde{\eta}_{12}$ or negative values
of the diagonal elements of the turbulent diffusion tensor
favour growth at early times. Preliminary work on simple
models of velocity dynamics that we are exploring suggests that
$\tilde{\eta}_{22}$ can become negative but $\tilde{\eta}_{11}$
and $\tilde{\eta}_{33}$ remain positive; this happens because the turbulence is strongly 
affected by the background shear and 
the velocity correlators are not isotropic.
Thus a non-zero $k_2$ seems to be required for
growth initially.

At intermediate times, when the $t^2$ term dominates we can always choose
shearing waves with an appropriate sign and magnitude of $k_1k_2$ such that 
$2At^2 (\tilde{\eta}_{11}k_1k_2 + \tilde{\eta}_{12}k_2^2)$ is negative, 
and there is growth of the mean field.
On the other hand, all shearing waves with non-zero $k_2$ will eventually decay,
in the long time limit $t\to \infty$, 
if $\tilde{\eta}_{11} > 0$, as then the
$t^3$ term is negative definite.
Thus it seems likely that the shear dynamo can have
shearing wave solutions which grow for some time if
they have non-zero $X_2$ dependence, but which will eventually decay.
As already emphasized above, one needs to develop a dynamical theory
of the velocity correlators, for deriving more explicit results on dynamo action, due to non--helical
turbulence and shear.  It is, in general, not an easy task to make analytical progress on a
dynamical theory. However, in the limit of low fluid Reynolds numbers, a
perturbative analysis is possible and the velocity correlators can be
computed explicitly. Such an analysis has been undertaken by Singh and   
Sridhar, and preliminary results for non--helical forcing indicate that
the turbulent diffusion coefficient $\tilde{\eta}_{22}$ can indeed become
negative. Also our conclusions are based on the differential equation
approximation, which is valid for a finite period and 
thus we need to solve the integral equation for the mean field
evolution directly, to firm up the above results.
 
\section{Conclusions}
We have studied here large--scale dynamo action due to turbulence 
in the presence of a linear shear flow.
Systematic use of the shearing coordinate transformation
and the Galilean invariance of a linear shear flow allows us to develop
a quasilinear theory of the shear dynamo which,
we emphasize, is non perturbative in the
shear parameter. The result is an integro--differential equation 
for the evolution of the mean magnetic field.
We showed using this equation that for
non helical turbulence, the time evolution of the cross--shear
components of the mean field do not depend on any other components 
excepting themselves. This implies that there is essentially no shear--current type effect  in 
quasilinear theory in the limit of zero resistivity. Our result is 
valid for any Galilean--invariant velocity field, independent of its dynamics.

We then derived differential equations for the mean-field evolution, 
by developing a systematic approximation to the integro-differential
equation, assuming the mean field varies on time scales much longer
than the correlation times of the turbulence.
For non-helical velocity correlators, 
these equations can be solved in terms of shearing waves. These waves
can grow transiently at early and intermediate times.
However it is likely that they will eventually decay at asymptotically late times.
More explicit statements about the behaviour of the shearing wave
solutions requires developing a dynamical theory of velocity correlators 
in shear flows. It is also important to directly solve the integral equation for the mean field
as the differential equation approximation is valid only for a limited period.

Growth of large--scale magnetic fields in the presence of shear and non--helical turbulence has 
been reported in some direct numerical simulations \cite{BRRK08,Yousef08}. Whether we can 
understand these numerical results through our quasilinear theory depends on the existence (or otherwise)
of growing solutions to the integral equation~(\ref{Hinduction}) for the mean field. This in turn relies on the form 
of the velocity correlators, which will be strongly affected by shear and highly anisotropic; hence it is difficult to guess their 
tensorial forms {\it a priori}, and it is necessary to develop a dynamical theory of velocity correlators. We cannot discount the 
possibility that effects we have ignored may also play a role. Perhaps the initial growth of the shearing wave in the mean field, for
large enough shear, is sustained by an effect which breaks one
of our assumptions. One possibility is that  
helicity fluxes arising due to shear, turbulence and an inhomogeneous
mean magnetic field \cite{VC01,BS05} 
induce a nonlinear alpha effect when the Lorentz forces become
strong. Another is the possible presence of an 
incoherent alpha-shear dynamo 
\cite{VB97,BRRK08} in these simulations. A third possibility is that if 
even transient growth makes non--axisymmetric mean fields
strong enough, they themselves might drive motions which could lead to  sustained dynamo action; 
this seems remniscent of some of the subcritical dynamos discussed by \cite{rin08}.
Clearly further studies of various aspects of the shear dynamo,
particularly incorporating velocity dynamics can only
be more fruitful.

\begin{acknowledgments}
We acknowledge Nordita for providing a stimulating atmosphere
during the program on `Turbulence and Dynamos'.
We thank Axel Brandenburg, Karl-Heinz R\"adler and
Matthias Rheinhardt for valuable comments.
\end{acknowledgments}

\end{document}